\documentstyle[11pt]{article}
\begin{document}
\renewcommand{\refname}{References.}
\newcommand{\Nf}{N_{\!f}}
\newcommand{\Nc}{N_{\!c}}
\newcommand{\partialslash}{\partial \!\!\! /}
\newcommand{\kslash}{k \!\!\! /}
\newcommand{\half}{{\mbox{\small{$\frac{1}{2}$}}}}
\title{Quark, gluon and ghost anomalous dimensions at $O(1/N_f)$ in
quantum chromodynamics.}
\author{J.A. Gracey, \\ Department of Applied Mathematics and Theoretical
Physics, \\ University of Liverpool, \\ P.O. Box 147, \\ Liverpool, \\ L69 3BX,
\\ United Kingdom.}
\date{}
\maketitle
\vspace{5cm}
\noindent
{\bf Abstract.} By considering the scaling behaviour of various Feynman graphs
at leading order in large $\Nf$ at the non-trivial fixed point of the
$d$-dimensional $\beta$-function of QCD we deduce the critical exponents
corresponding to the quark, gluon and ghost anomalous dimensions as well as the
anomalous dimensions of the quark-quark-gluon and ghost-ghost-gluon vertices in
the Landau gauge. As the exponents encode all orders information on the
perturbation series of the corresponding renormalization group functions we
find agreement with the known three loop structure and, moreover, we provide
new information at all subsequent orders.

\vspace{-18cm}
\hspace{10cm}
{\bf LTH-319}
\newpage
One of the more difficult exercises in understanding the quantum field
theories, such as quantum chromodynamics, which underpin the dynamics of the
particles we observe in nature, is in gaining knowledge of their perturbative
structure to very high orders. For example, following the early one and two
loop work of \cite{1,2,3} the renormalization group functions of QCD are only
known to third order, \cite{4,5}. To a large degree one is hindered from
pushing such calculations to a further order, which will be necessary due to
the increase in experimental energies envisaged at LHC and the SSC, by the huge
numbers of Feynman diagrams one would have to compute, a subclass of which
possess an intricate integral structure. Indeed such calculations can only be
tackled by ingenious algorithms which are executed on computer. Whilst the
basic functions of the renormalization group equation are ordinarily calculated
at successive orders in the perturbative coupling constant, which we will
denote by $g$, it is not the only parameter with which one can compute with.
For theories, such as QCD which possess $\Nf$ flavours of quarks one can also
expand the functions in powers of $1/\Nf$ and compute various coefficients. The
technique to achieve this was introduced in \cite{6,7} for the $O(N)$ bosonic
$\sigma$ model and later extended to the $\sigma$ model on $CP(N)$, which is a
$U(1)$ gauge theory, and then to QED, \cite{9,10}. Basically, it entails
computing various critical exponents at the $d$-dimensional fixed point of the
theory, defined as the non-trivial zero of the $\beta$-function, where the
theory has a conformal symmetry, \cite{6}. The exponents which emerge depend on
the space-time dimension, $d$, and are computed order by order in powers of
$1/\Nf$. Further, as they are calculated at a fixed point the renormalization
group equation takes a simpler form there to the extent that exponents have
simple relations with the corresponding renormalization group equations, such
as the wave function or vertex renormalization, at criticality, \cite{11}.
Moreover, one can undo such relations to deduce information on the perturbative
structure at non-critical $g$. In particular the coefficients of each power of
the coupling constant, at the order in large $N$ one is working to, can be
determined and they correspond to those of the $\overline{\mbox{MS}}$ scheme.
This can be understood from the fact that at the critical point there is a
conformal symmetry and thus the exponents must correspond to a mass independent
scheme. Another success of the method of \cite{6,7} is the possibility of
pushing the calculations to $O(1/\Nf^2)$, \cite{10}, which is a depth not
possible by conventional large $N$ methods.

The purpose of this letter is to set up the analogous approach for QCD,
extending the earlier QED work of \cite{9}. The aim is to deduce non-trivial
exponents for various fields in QCD and demonstrate their agreement with the
known three loop perturbative $\overline{\mbox{MS}}$ calculations, \cite{4,5},
and then to deduce information at higher orders. By non-trivial we mean
determining exponents which are not merely deduced by examining the analogous
QED results and multiplying those by the appropriate non-abelian group factors.
Such non-trivial results will be derived either from graphs involving ghost
fields, the triple gluon vertex or those which would otherwise be absent in QED
by Furry's theorem. In particular, we will deduce the anomalous dimensions of
the quark, gluon and ghost fields at leading order in large $\Nf$ as well as
the vertex anomalous dimensions of the quark-quark-gluon (qqg) and
ghost-ghost-gluon (gcc) vertices, in the Landau gauge. Although these field
anomalous dimensions are gauge dependent they must be computed initially before
attempting to determine gauge independent quantities such as the
$\beta$-function which will be a considerably more involved computation and is
our long term goal. This is akin to the perturbative approach when the wave
function renormalization of each field must be determined first.

The version of the QCD lagrangian we use to compute exponents is
\begin{eqnarray}
L &=& - \, \frac{(F^a_{\mu\nu})^2}{4e^2} + i \bar{\psi}^{iI}\partialslash
\psi^{iI} + A^a_\mu\bar{\psi}^{iI}T^a_{IJ}\gamma^\mu\psi^{iJ}
- \frac{1}{e^2} f^{abc} \partial_\mu A^a_\nu A^{\mu b} A^{\nu c}
\nonumber \\
&&-~ \frac{1}{4e^2} f^{abc} f^{ade} A^b_\mu A^c_\nu A^{\mu d} A^{\nu e}
- \frac{1}{2\xi e^2} (\partial_\mu A^{\mu a})^2 \nonumber \\
&&-~ \partial^\mu \bar{c}^a \partial_\mu c^a + f^{abc} \partial_\mu \bar{c}^a
c^b A^{\mu c}
\end{eqnarray}
where $\psi^{iI}$ is the quark field, $1$ $\leq$ $i$ $\leq$ $\Nf$, $1$ $\leq$
$I$ $\leq$ $\Nc$, $A^a_\mu$ is the gluon field, $1$ $\leq$ $a$ $\leq$ $N^2_c$
$-$ $1$, $F^a_{\mu\nu}$ $=$ $\partial_\mu A^a_\nu$ $-$ $\partial_\nu A^a_\mu$,
$c^a$ and $\bar{c}^a$ are the ghost fields, $\xi$ is the covariant gauge
parameter, $T^a_{IJ}$ are the generators of the non-abelian gauge group,
$f^{abc}$ its structure constants and $e$ is the coupling constant. It is
important to note that we have absorbed a power of the coupling constant into
the definition of the gauge field so that the qqg vertex has the same style of
interaction as the electron photon vertex of QED, \cite{9,10}, which was
primarily for applying the techniques of uniqueness of \cite{12} to computing
massless Feynman diagrams which arise at $O(1/\Nf^2)$. To make more explicit
the notions already discussed we note that the fixed point where we will
concentrate our analysis is given by $\beta(g_c)$ $=$ $0$, where $g_c$ is the
non-trivial zero of the $d$ $=$ $4$ $-$ $2\epsilon$ dimensional
$\beta$-function. It has been calculated in $\overline{\mbox{MS}}$ using
dimensional regularization and in terms of the dimensionless coupling
constant $g$ $=$ $(e/2\pi)^2$, in the notation of \cite{13}, it is, [1-5],
\begin{eqnarray}
\beta(g) &=& (d-4)g + \left[ \frac{2}{3}T(R)\Nf - \frac{11}{6}C_2(G) \right]
g^2 \nonumber \\
&+& \left[ \frac{1}{2}C_2(R)T(R)\Nf + \frac{5}{6}C_2(G)T(R)\Nf
- \frac{17}{12}C^2_2(G) \right] g^3 \nonumber \\
&-& \left[ \frac{11}{72} C_2(R)T^2(R)\Nf^2 + \frac{79}{432} C_2(G) T^2(R)
\Nf^2 \right. \nonumber \\
&&+~ \left. \frac{1}{16} C^2_2(R) T(R) \Nf - \frac{205}{288}C_2(R)C_2(G)T(R)\Nf
\right. \nonumber \\
&&-~ \left. \frac{1415}{864} C^2_2(G)T(R)\Nf + \frac{2857}{1728}C^3_2(G)
\right] g^4 + O(g^5)
\end{eqnarray}
from which it follows that
\begin{eqnarray}
g_c &=& \frac{3\epsilon}{T(R)\Nf} + \frac{1}{4T^2(R)\Nf^2} \left[ \frac{}{}
33C_2(G)\epsilon - \left( 27C_2(R) + 45C_2(G)\right) \epsilon^2 \right.
\nonumber \\
&+& \left. \left( \frac{99}{4}C_2(R) + \frac{237}{8} C_2(G) \right)
\epsilon^3 + O(\epsilon^4) \right] + O \left( \frac{1}{\Nf^3} \right)
\end{eqnarray}
The Casimirs for a general classical Lie group which appear in (2) and (3) are
defined via
\begin{equation}
\mbox{Tr}(T^aT^b) ~=~ T(R) \delta^{ab} ~~,~~ T^aT^a ~=~ C_2(R) I ~~,~~
f^{acd} f^{bcd} ~=~ C_2(G) \delta^{ab}
\end{equation}
and for $SU(\Nc)$, $T(R)$ $=$ $\half$, $C_2(R)$ $=$ $(N^2_c-1)/2N_c$ and
$C_2(G)$ $=$ $\Nc$.

At $g_c$ the fields of (1) obey simple power law behaviour in analogy with
ideas in statistical mechanics and therefore in momentum space the asymptotic
scaling forms of the propagators of each field in the critical region as $k^2$
$\rightarrow$ $\infty$ are, \cite{9},
\begin{eqnarray}
\psi(k) &\sim& \frac{A\kslash}{(k^2)^{\mu-\alpha}} ~~,~~
A_{\mu\nu} ~\sim~ \frac{B}{(k^2)^{\mu-\beta}}\left[ \eta_{\mu\nu}
- \frac{k_\mu k_\nu}{k^2} \right] \nonumber \\
c(k) &\sim& \frac{C}{(k^2)^{\mu-\gamma}}
\end{eqnarray}
The quantities $A$, $B$ and $C$ are the respective $k$-independent amplitudes
of each field and $\alpha$, $\beta$ and $\gamma$ are their dimensions which are
built out of the canonical (or classical) dimension and the anomalous
contribution. The latter arises as a result of quantum corrections such as
radiative corrections to the underlying Green's functions. In particular, we
define the exponents to be
\begin{equation}
\alpha ~=~ \mu - 1 + \half \eta ~~,~~ \beta ~=~ 1 - \eta - \chi ~~,~~
\gamma ~=~ \mu - 1 + \half \eta_c
\end{equation}
where $d$ $=$ $2\mu$ and $\eta$, $\eta_c$ and $\chi$ are respectively the
quark, ghost and qqg vertex anomalous dimensions. Each is $O(1/\Nf)$ and
depends on $\mu$ and $\Nc$ through the Casimirs of (4) and we will determine
their leading order large $\Nf$ structure here. Also their relation to the
conventional renormalization group functions are, \cite{13},
\begin{eqnarray}
\eta ~\longleftrightarrow~ \gamma_2(g_c) &,& \eta + \chi ~\longleftrightarrow~
\gamma_3(g_c) \nonumber \\
\chi ~\longleftrightarrow~ \gamma_1(g_c) &,& \eta_c ~\longleftrightarrow~
\tilde{\gamma}_3(g_c)
\end{eqnarray}
where $\gamma_2(g)$ is the quark wave function renormalization, $\gamma_1(g)$
is the quark gluon vertex renormalization function, $\gamma_3(g)$ is the gluon
wave function renormalization and $\tilde{\gamma}_3(g)$ corresponds to the
ghost wave function renormalization. The third relation of (6) arises from
analysing the ghost kinetic term in (1) but it can be related to other
anomalous dimensions by examining the ghost interaction term to obtain
\begin{equation}
\eta_c ~=~ \eta ~+~ \chi ~-~ \chi_c
\end{equation}
where $\chi_c$ is the gcc vertex anomalous dimension being defined in a similar
way to $\chi$. This result, which we will use later to deduce $\eta_c$ and
demonstrate its agreement with perturbation theory, is nothing more than an
expression of one of the Slavnov Taylor identities for QCD in exponent
language, \cite{14}. A similar observation concerning the QED Ward identity was
made in \cite{15}. As we are working in a gauge theory we must choose a
particular gauge to calculate within. As the covariant gauge parameter remains
unrenormalized in the Landau gauge, we have made the choice $\xi$ $=$ $0$,
\cite{9}. This is important for comparing the structure of gauge dependent
exponents with perturbation theory which we do later.

The method to solve for (7) involves examining the Dyson equations of various
Green's functions truncated to $O(1/\Nf)$ in the critical region where the
propagators of the Feynman graphs are replaced by the expressions (5). Although
the analysis of the quark and gluon $2$-point functions is trivial in the sense
defined earlier, it is worth discussing it to illustrate the simplicity of the
method and the Dyson equations are shown in fig. 1 where dotted lines
correspond to quarks and solid lines to ghosts. We use dressed propagators
since the effect of including a non-zero anomalous dimension in (5) is to
reproduce the infinite chain of bubble graphs one would ordinarily have to
consider, which was the original approach to examining QED in the large $\Nf$
expansion, \cite{16}. Using (5) means fewer Feynman graphs have to be analysed.
Several points concerning the graphs of fig. 1 ought to be noted. First, there
are no graphs in the gluon $2$-point function involving ghost loops, triple or
quartic gluon vertices since these are $O(1/\Nf)$ with respect to the quark
loop. (A similar set of graphs were analysed in a different context in
\cite{17}.) Second, the quantities $\psi^{-1}$, $A^{-1}_{\mu\nu}$ and $c^{-1}$
denote the respective $2$-point functions and their asymptotic scaling forms
can be deduced from (5) by simple momentum space inversion where the gluon is
inverted on the transverse subspace, \cite{9}, as
\begin{eqnarray}
\psi^{-1}(k) &\sim& \frac{\kslash}{A(k^2)^{\alpha-\mu+1}} ~~,~~
A^{-1}_{\mu\nu}(k) ~\sim~ \frac{1}{B(k^2)^{\beta-\mu}} \left[ \eta_{\mu\nu}
- \frac{k_\mu k_\nu}{k^2} \right] \nonumber \\
c^{-1}(k) &\sim& \frac{1}{C(k^2)^{\gamma-\mu}}
\end{eqnarray}
Thus the first two graphs of fig. 1 can be represented in the critical region
by
\begin{eqnarray}
0 &=& 1 + \frac{(2\mu-1)(\beta+1-\mu)\alpha}{(\mu-\beta)(\alpha+\beta)}
a(\mu-\beta)a(\mu-\alpha)a(\alpha+\beta) C_2(R) z \\
0 &=& 1 + \frac{8\alpha^3T(R)\Nf}{(2\alpha+1)} a^2(\mu-\alpha)a(2\alpha+1) z
\end{eqnarray}
where only the transverse part of the gluon equation is physically relevant,
\cite{9}, $\mbox{tr}1$ $=$ $4$, $z$ $=$ $A^2B$ (and $y$ $=$ $BC^2$) and we have
set $a(\alpha)$ $=$ $\Gamma(\mu-\alpha)/\Gamma(\alpha)$ for all $\alpha$. In
(10) and (11) the only unknowns at leading order in $1/\Nf$ are $\eta_1$ and
$z_1$, where for example $\eta$ $=$ $\sum_{i=1}^\infty \eta_i/N^i_{\! f}$ with
$\eta_i$ $=$ $\eta_i(\mu,\Nc)$. Although it may appear that $\chi_1$ also
arises through the factor $a(\alpha+\beta)$ in (10), in the momentum space
approach to solving the Dyson equations in this critical point method, one
sets $\chi_1$ $=$ $0$ at leading order so that $\beta$ $=$ $1$ $-$ $\eta$ here.
This is partly to ensure consistency between the momentum and coordinate space
solutions for $\eta_1$, which can easily be verified by examining the $x$-space
QED solution, \cite{10}, or considering the same calculation in the toy $O(N)$
$\sigma$ model of \cite{18}. More importantly, though, since the method we are
using relies very much on the conformal nature of the fixed point, the qqg
$3$-point vertex is in fact only conformal or unique, \cite{10}, at leading
order when $\chi_1$ $=$ $0$, \cite{6,7,10}, which justifies this point.
Further, the factors $(p^2)^{-\chi}$ have been omitted from the second terms of
(10) and (11) partly for this reason but also because $\chi_1$ $=$ $O(1/\Nf)$
and so its $1/\Nf$ expansion, ie $1$ $-$ $(\chi_1\ln p^2)/\Nf$, means it will
only be important for deducing $\eta_2$. The treatment of this can be seen, for
example, in the QED work of \cite{10}. Thus eliminating $z_1$ between (10) and
(11) yields
\begin{equation}
\eta_1 ~=~ \frac{(2\mu-1)(\mu-2)\Gamma(2\mu) C_2(R)}{4\Gamma^2(\mu)
\Gamma(\mu+1) \Gamma(2-\mu) T(R)}
\end{equation}
which corresponds to the QED result of \cite{9} aside from the non-abelian
factor of $C_2(R)/T(R)$ $=$ $8/3$ for QCD and agrees with the $3$-loop
perturbative expansion of $\gamma_2(g_c)$.

To deduce the vertex anomalous dimensions one carries out a similar analysis
using instead the truncated leading order large $\Nf$ graphs contributing to
the qqg and gcc $3$-point functions which are illustrated in figs 2 and 3
respectively. However, naively computing each individual graph with the leading
order exponents one would discover that they are infinite and therefore need to
be regularized. This is achieved by shifting the gluon exponent by an
infinitesimal amount $\Delta$, $\beta$ $\rightarrow$ $\beta$ $-$ $\Delta$, with
$\Delta$ playing a similar role to the $\epsilon$ in dimensional
regularization, \cite{10}. The aim now is, following the ideas of \cite{18}, to
compute each graph with the shifted exponent to observe that they have the
following structure at $O(1/\Nf)$
\begin{equation}
\frac{P}{\Delta} ~+~ Q ~+~ R \ln p^2 ~+~ O(\Delta)
\end{equation}
where $P$, $Q$ and $R$ will depend on $\mu$ and $\Nc$ and $p$ is the external
momentum flowing through the external quark or ghost fields. The simple pole
of (13) is absorbed into the usual vertex counterterm, \cite{18}. This leaves
a $\Delta$-finite Green's function but the scaling behaviour would be
spoiled by the presence of the $\ln p^2$ term. However, general arguments,
\cite{8,18}, imply that the overall scaling behaviour of the qqg vertex, for
instance, has to be $(p^2)^{\chi/2}$ and therefore the $\ln p^2$ terms at
each order in $1/\Nf$ must re-sum to this form which allows one to determine
an expression for $\chi$.

Thus substituting (5) into the graphs of fig. 2 the first graph involves
multiplying the QED result of \cite{15} by the appropriate group factor,
so that it contributes
\begin{equation}
- \, \left[ C_2(R) - \frac{C_2(G)}{2} \right] \frac{\eta^{\mbox{o}}_1}{T(R)}
\end{equation}
to $\chi_1$ where $\eta_1$ $\equiv$ $C_2(R) \eta^{\mbox{o}}_1/T(R)$. Useful in
computing the contribution from the two loop graphs of fig. 2 is the result
\begin{eqnarray}
\!\! \int_k \frac{k_\mu k_\nu}{(k^2)^\alpha ((k-p)^2)^\beta} &=&
\!\frac{(\alpha+\beta-\mu-2)a(\alpha-1)a(\beta-1)a(2\mu-\alpha-\beta+2)}
{2(\alpha-1)(\beta-1)(p^2)^{\alpha+\beta-\mu-1}} \nonumber \\
&&\times \left[ \eta_{\mu\nu} + \frac{2p_\mu p_\nu}{p^2} \frac{(\mu-\alpha+1)
(\alpha+\beta-\mu-1)}{(\mu-\beta)} \right]
\end{eqnarray}
valid for all $\alpha$ and $\beta$ and they and the second graph of fig. 2 are
proportional to $C_2(G)$ and contribute
\begin{equation}
- \, \frac{(\mu-1)C_2(G)\eta^{\mbox{o}}_1}{2(\mu-2)T(R)}
\end{equation}
to $\chi_1$. Thus, we have
\begin{equation}
\chi_1 ~=~ - \, \left[ C_2(R) + \frac{C_2(G)}{2(\mu-2)} \right]
\frac{\eta^{\mbox{o}}_1}{T(R)}
\end{equation}
from which we deduce the gluon anomalous dimension as
\begin{equation}
\eta_1 \, + \, \chi_1 ~=~ - \, \frac{C_2(G) \eta^{\mbox{o}}_1}{2(\mu-2)T(R)}
\end{equation}
To check that (17) and (18) are in agreement with perturbation theory we recall
that the three loop $\overline{\mbox{MS}}$ result of \cite{5} in arbitrary
covariant gauge, in our notation, is
\begin{eqnarray}
\gamma_1(g) &=& \left[ \xi C_2(R) + \frac{(3+\xi)}{4}C_2(G)\right] \frac{g}{2}
\nonumber \\
&+& \left[ \frac{17}{2}C_2(R)C_2(G) - \frac{3}{2}C_2^2(R) + \frac{67}{24}
C^2_2(G) - 2C_2(R)T(R)\Nf \right. \nonumber \\
&&-~ \left. \frac{5}{6} C_2(G) T(R)\Nf - (1-\xi)C_2(G)\left( \frac{5}{2} C_2(R)
+ \frac{15}{16}C_2(G) \right) \right. \nonumber \\
&&+~ \left. \frac{(1-\xi)^2C_2(G)}{4} \left( C_2(R) + \frac{1}{2}C_2(G)
\right) \right] \frac{g^2}{8} \nonumber \\
&+& \left[ \frac{3}{2}C^3_2(R) + C^2_2(R)C_2(G) \left( 12 \zeta(3)
- \frac{143}{4} \right) \right. \nonumber \\
&&+~ \left. C_2(R)C_2^2(G) \left( \frac{10559}{144} - \frac{15}{2} \zeta(3)
\right) + \left( \frac{3}{4}\zeta(3) + \frac{10703}{864} \right) C^3_2(G)
\right. \nonumber \\
&&+~ \left. 3C^2_2(R)T(R)\Nf + 2C_2(R)C_2(G)T(R)\Nf \left( 6\zeta(3)
- \frac{853}{36} \right) \right. \nonumber \\
&&-~ \left. C^2_2(G)T(R)\Nf \left( 9\zeta(3) + \frac{205}{108} \right)
+ \frac{20}{9} C_2(R) T^2(R)\Nf^2 \right. \nonumber \\
&&-~ \left. \frac{35}{27} C_2(G) T^2(R) \Nf^2
- (1-\xi)\left( C_2(R)C^2_2(G) \left( \frac{3}{2}\zeta(3)
+ \frac{371}{32} \right) \right. \right. \nonumber \\
&&+~ \left. \left. \frac{C^3_2(G)}{32} \left( 18\zeta(3) + 127 \right)
-  C_2(G)T(R)\Nf \!\left(\! \frac{17}{4}C_2(R) + C_2(G) \!\right)
\!\right) \right. \nonumber \\
&&+~ \left. (1-\xi)^2 C_2^2(G)\!\left(\! C_2(R) \! \left(\frac{3}{8}\zeta(3)
+ \frac{69}{32} \right) + \frac{C_2(G)}{32} \left( 3\zeta(3) + 27 \right)
\right) \right. \nonumber \\
&&-~ \left. (1-\xi)^3 C^2_2(G)\left( \frac{5}{16}C_2(R) + \frac{7}{64}
C_2(G) \right) \right] \frac{g^3}{32}
\end{eqnarray}
which with (3) gives, in the Landau gauge,
\begin{eqnarray}
\gamma_1(g_c) &=& \frac{C_2(G)}{T(R)\Nf} \left[ \frac{9}{8} \epsilon
- \frac{15}{16}\epsilon^2 - \frac{35}{32}\epsilon^3 + O(\epsilon^4) \right]
\nonumber \\
&-& \frac{C_2(R)}{T(R)\Nf} \left[ \frac{9}{4}\epsilon^2 - \frac{15}{8}
\epsilon^3 + O(\epsilon^4) \right] + O \left( \frac{1}{\Nf^2} \right)
\end{eqnarray}
Expanding (17) in powers of $\epsilon$ $=$ $2$ $-$ $\mu$ to $O(\epsilon^3)$, it
is easy to see that it is in complete agreement with (20) which is a stringent
check on our results. More importantly, the higher $O(1/\Nf)$ coefficients of
(19) can now be deduced by extending this argument. If we write the leading
order part of (19) as
\begin{equation}
\gamma_1(g) ~=~ \frac{3}{8} C_2(G) g + \sum_{n=2}^\infty a_n (T(R)\Nf)^{n-1}
g^n
\end{equation}
then (17) implies
\begin{eqnarray}
a_4 &=& \frac{1}{36} \left[ \frac{35}{36}C_2(R) + \left( \zeta(3)
- \frac{83}{144}\right) C_2(G)\right] \nonumber \\
a_5 &=& \! - \, \frac{1}{108} \! \left[ C_2(R) \left( \! 2\zeta(3)
- \frac{83}{72} \right) - \frac{C_2(G)}{2} \! \left( \! 3\zeta(4)
- \frac{5\zeta(3)}{3} - \frac{65}{48} \! \right) \right]
\end{eqnarray}
It is interesting to note that the first appearance of the $\xi$ parameter in
(19) and also in the other three loop renormalization group functions of
\cite{5} is at $O(1/N^2_{\!f})$. In fact this is true to all orders so that the
leading order large $\Nf$ coefficients are gauge independent. Thus the
exponents we have written down encode gauge independent information, at all
orders, of part of the renormalization group functions.

Repeating the analysis for the gcc vertex graphs of fig. 3, one finds that each
is $\Delta$-finite and therefore at $O(1/\Nf)$
\begin{equation}
\chi_{c \, 1} ~=~ 0
\end{equation}
which implies from (8) that
\begin{equation}
\eta_{c \, 1} ~=~ - \, \frac{C_2(G)\eta^{\mbox{o}}_1}{2(\mu-2)T(R)}
\end{equation}
Expanding (24) in powers of $\epsilon$ as before and comparing with the two
loop arbitrary gauge calculation of \cite{3} one again finds agreement. The
three loop ghost anomalous dimension has been calculated in \cite{4}. Although
that calculation was carried out in the Feynman gauge the three loop $O(1/\Nf)$
coefficient is in agreement with the $O(\epsilon^3)$ term of the expansion of
(24) in the Landau gauge and our observation concerning the non-appearance of
$\xi$ at third order at $O(1/\Nf)$ in previous renormalization group functions
also  hold at this order.

Finally, it is worth recording the leading order form of a gauge independent
exponent which encodes all orders coefficients on the perturbative structure of
the renormalization of the quark mass, $\gamma_m(g)$. It is trivial in the
sense we mentioned earlier and is deduced by computing the anomalous dimension
of the composite operator $\bar{\psi}\psi$ within the Green's function $\langle
\psi [\bar{\psi}\psi] \bar{\psi} \rangle$. This was discussed originally in
\cite{16} for QED and including the structure which arises from the non-abelian
nature of (1), we find that the exponent is
\begin{equation}
\gamma_m(g_c) ~=~ - \, \frac{2C_2(R) \eta^{\mbox{o}}_1}{(\mu-2)T(R)\Nf}
\end{equation}
and its $\epsilon$-expansion is in agreement with the $\overline{\mbox{MS}}$
perturbation series of \cite{19}.

We conclude by remarking that we have provided the first stage in the analysis
of QCD in the large $\Nf$ expansion where the aim is to deduce the coefficients
of the perturbative renormalization group functions in order to provide new
results as well as checking existing results. It ought now to be possible to
deduce the $\beta$-function in the same approximation as well as building on
the $O(1/N^2_{\! f})$ QED work of \cite{10} to gain a deeper insight into the
theory.

\vspace{1cm}
\noindent
{\bf Acknowledgement.} The author thanks Drs D.J. Broadhurst and D.R.T. Jones
for useful discussions.

\newpage
\noindent
{\Large {\bf Figure Captions.}}
\begin{description}
\item[Fig. 1.] Leading order skeleton Dyson equations where dotted lines are
quarks and solid lines ghosts.
\item[Fig. 2.] Leading order graphs for $A^a_\mu\bar{\psi}\gamma^\mu T^a\psi$.
\item[Fig. 3.] Leading order graphs for $f^{abc}A^a_\mu\partial^\mu
\bar{c}^bc^c$.
\end{description}
\end{document}